\documentclass[aps,prb,twocolumn,superscriptaddress,showpacs]{revtex4}
\usepackage{graphicx}
\usepackage{dcolumn} 
\usepackage{bm}      
\usepackage{amsmath,amssymb,times}
\usepackage{color}
\usepackage{xcolor}

\begin{document}

\title{Magnetic transitions and magnetodielectric effect in the antiferromagnet SrNdFeO$_4$}

\author{J.~M.~Hwang},
\affiliation{National High Magnetic
Field Laboratory, Florida State University, Tallahassee, Florida
32310-3706, USA}
\affiliation{Department of Physics, Florida State University,
Tallahassee, Florida 32306-3016, USA}

\author{E.~S.~Choi},
\affiliation{National High Magnetic
Field Laboratory, Florida State University, Tallahassee, Florida
32310-3706, USA}

\author{H.~D.~Zhou},
\affiliation{National High Magnetic
Field Laboratory, Florida State University, Tallahassee, Florida
32310-3706, USA}

\author{Y.~Xin},
\affiliation{National High Magnetic
Field Laboratory, Florida State University, Tallahassee, Florida
32310-3706, USA}

\author{J.~Lu},
\affiliation{National High Magnetic
Field Laboratory, Florida State University, Tallahassee, Florida
32310-3706, USA}

\author{P.~Schlottmann}
\affiliation{National High Magnetic
Field Laboratory, Florida State University, Tallahassee, Florida
32310-3706, USA}
\affiliation{Department of Physics, Florida State University,
Tallahassee, Florida 32306-3016, USA}

\date{\today}

\begin{abstract}
We investigated the magnetic phase diagram of single crystals of SrNdFeO$_{4}$ by measuring the magnetic properties, the specific heat and the dielectric permittivity. The system has two magnetically active ions, Fe$^{3+}$ and Nd$^{3+}$. The Fe$^{3+}$ spins are antiferromagnetically ordered below 360 K with the moments lying in the $ab$-plane, and undergo a reorientation transition at about 35-37 K to an antiferromagnetic order with the moments along the $c$-axis. A short-range, antiferromagnetic ordering of Nd$^{3+}$ along the $c$-axis was attributed to the reorientation of Fe$^{3+}$ followed by a long-range ordering at lower temperature [S. Oyama {\it et al.} J. Phys.: Condens. Matter. {\bf 16}, 1823 (2004)]. At low temperatures and magnetic fields above 8 T, the Nd$^{3+}$ moments are completely spin-polarized. The dielectric permittivity also shows anomalies associated with spin configuration changes, indicating that this compound has considerable coupling between spin and lattice. A possible magnetic structure is proposed to explain the results.
\end{abstract}


\maketitle

\section{Introduction}

\begin{figure}[tbp]
\linespread{1}
\includegraphics[width=2.8in]{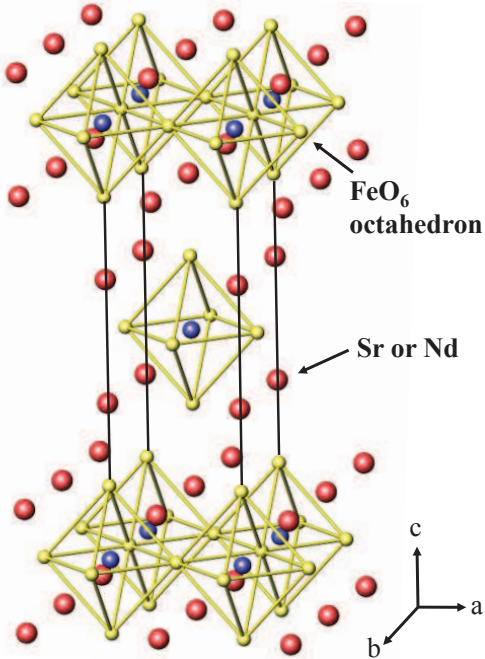}
\caption{The crystal structure of SrNdFeO$_4$. Red spheres are Sr or Nd atoms and the octahedra are FeO$_6$ units (Fe atoms are blue spheres).}
\end{figure}

Magnetic systems involving more than one species of magnetically active ions are always a subject of interest. For example, the high performance permanent magnets are utilizing the interaction between transition metal ions and rare-earth ions to achieve high coercivity in intermetallic compounds.\cite{Buschow,Hummler} Some multiferroic materials also possess two magnetic ions which play a crucial role in the realization of the multiferroicity via non-collinear spin structures or magnetostriction.\cite{Chapon,Tokunaga,Kimura_TbMnO3,Kenzelmann,Arima,Hur}

In the present case SrNdFeO$_{4}$ has two kinds of magnetic ions, namely, Fe$^{3+}$ and Nd$^{3+}$. SrNdFeO$_4$ crystallizes in the tetragonal K$_2$NiF$_4$-type structure with space group $I4/mmm$ and corresponds to the $n$ = 1 compound of the Ruddlesden-Popper series with general formula A$_{n+1}$B$_n$O$_{3n+1}$. Here A corresponds to either Sr or Nd ions, and B represents Fe ions. In this structure, perovskite ABO$_3$ and rock salt AO layers alternate along the tetragonal $c$-axis. Sr$^{2+}$ and Nd$^{3+}$ ions are randomly distributed at the same Wyckoff 4e positions as shown in Fig.~1.\cite{Oyama} Layered perovskite structures have been widely investigated for their strong magneto-electric coupling and the possibility of finding new multiferroic materials.\cite{Komskii,Eerenstein,Cheong}

The $^{57}$Fe M\"{o}ssbauer spectra and neutron scattering in polycrystalline samples of SrNdFeO$_4$ revealed that the Fe$^{3+}$ moments order antiferromagnetically below $T_N$ = 360 K with the moments lying in the $ab$-plane.\cite{Oyama} The Fe$^{3+}$ moments undergo a spin-reorientation transition from $ab$-plane to $c$-axis at about 36 K, which is accompanied by the onset of short-range ordering of Nd$^{3+}$ spins. It was also reported that a long-range antiferromagentic order along the $c$-axis of Nd$^{3+}$ spins occurs at $T_N$ = 15 K.

In this work, we have used high quality single crystal SrNdFeO$_4$ samples to study its magnetic phase transitions and magnetodielectric effect at high magnetic fields. We employed various experimental techniques to measure the temperature and magnetic field dependence of the magnetization, the specific heat and the dielectric permittivity. We verified the spin-reorientation transition of the Fe$^{3+}$ moments, the N\'eel transition of the Nd$^{3+}$ moments and found a first order spin-flop transition of Nd$^{3+}$ moments in high magnetic fields of the order of 9 T. The spin-flop transition is very close to the magnetic field where the complete spin polarization of the Nd$^{3+}$ occurs. Remarkably, the magnetic transitions are reflected in the dielectric permittivity, which suggests magnetoelectric and/or magnetoelastic effects are present in the title compound. We examine possible origins of the magnetodielectric effect considering the spin structure and the magnetic point group symmetry. Possible spin structures under magnetic fields are also proposed, which explain the experimental data.

\section{Experimental}

Single crystals of SrNdFeO$_4$ were grown by the traveling-solvent floating-zone (TSFZ) technique. The feed and seed rods for the crystal growth were prepared by solid state reaction. Stoichiometric mixtures of SrCO$_3$, Nd$_2$O$_3$ and Fe$_2$O$_3$ were ground together and pressed into 6-mm-diameter 60-mm rods under 400 atm hydrostatic pressure, and then calcined in Ar at 1200 $^{\circ}$C for 24 hours. The crystal growth was carried out in argon in an IR-heated image furnace (NEC model SC1-NDH) equipped with two halogen lamps and double ellipsoidal mirrors with feed and seed rods rotating in opposite directions at 25 rpm during crystal growth at a rate of 5 mm/h. X-ray Laue diffraction was used to confirm the crystal quality and orient the crystal.

The single crystal SrNdFeO$_4$ sample was also verified by transmission electron microscopy in a probe corrected JEM-ARM200F. The DC magnetization measurements were made with a vibrating sample magnetometer (VSM) of the Physical Property Measurement System (PPMS) by Quantum Design and with a high field VSM system of the National High Magnetic Field Laboratory for magnetic fields up to 35 T. The PPMS was also used to measure the specific heat. Thin plate-like single crystal samples were selected for the dielectric permittivity measurement. The electrodes were painted with silver epoxy on the two parallel opposite surfaces, whose normal direction is parallel to the $c$ or $a$-axis. The dielectric permittivity was measured using an Andeen-Hagerling AH-2700A capacitance bridge operating at a frequency of 10 kHz.

\section{Results}

\begin{figure}[tbp]
\linespread{1}
\par
\includegraphics[width=3.4in]{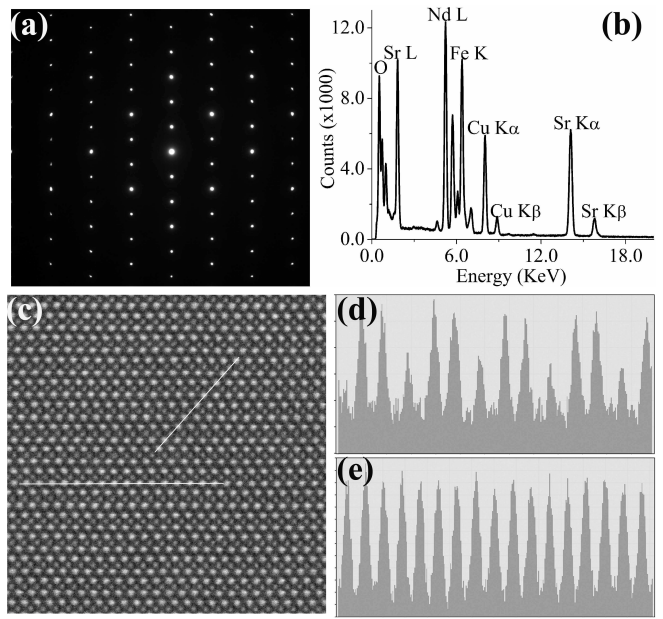}
\par
\caption{(a) Selected area diffraction pattern along [100] $a$-axis of the single crystal. (b) EDS spectrum of the crystal. (c) Atomic resolution STEM HAADF Z-contrast image looking down [100]. The Sr/Nd atomic columns show brighter contrast and the Fe atomic columns show weaker contrast. (d) Line profile of the atomic columns intensity indicated by the inclined line in (c). (e) Line profile of the atomic columns intensity indicated by the horizontal line in (c).}
\end{figure}

Fig. 2(a) illustrates a typical single crystal diffraction pattern of the [100] zone axis of the crystal. The energy dispersive x-ray spectrum collected from the sample confirms the stoichiometric composition containing Sr, Nd, Fe and O (Fig.~2(b)). Fig. 2(c) shows the atomic resolution image of the crystal down [100] by scanning transmission electron microscopy (STEM) high angle annular dark field (HAADF) Z-contrast imaging.  This type of image is sensitive to the atomic number Z, so that the intensity of atom columns is proportional to Z$^{n}$, where $n$ is close to 2.\cite{Pennycook, Loane} Therefore, the heavier atoms Sr/Nd show brighter contrast, while the Fe atoms are the smaller ones with weaker intensity. The line profile in Fig.~2(d) and (e) displays the intensity difference along the two directions (shown in Fig.~2(c)) more clearly. The intensity line profiles of Sr/Nd atomic columns exhibit similar intensity (Fig.~2(e)), which indicates that Sr and Nd atoms are uniformly mixed in the crystal.

\begin{figure}[tbp]
\includegraphics[width=3.4in]{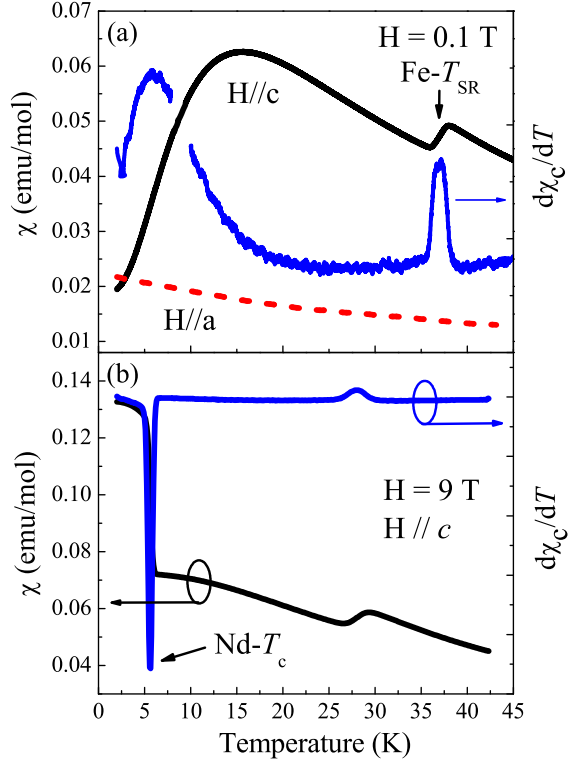}
\caption{$\chi$($T$) (left axis) (a) in $H$-field of 0.1 T applied along the $c$-axis (solid) and the $a$-axis (dashed), and (b) in $H$-field of 9 T applied along the $c$-axis. $d\chi_{c}/dT$ in $H$-field along the $c$-axis is plotted against the right axis.}
\end{figure}

The temperature dependence of the magnetization per mole divided by the external field ($\chi$($T$)) for single crystalline SrNdFeO$_{4}$ measured at $H$ = 0.1 T and 9 T is shown in Fig.~3(a) and 3(b), respectively. The  $\chi$($T$) agree qualitatively with the measurements on polycrystalline samples of Ref.~\onlinecite{Oyama}. As seen in Fig.~3(a), the $\chi$($T$) with the magnetic field along the $c$-axis shows two anomalies at 15 K and 38 K, respectively. These anomalies correspond well to the ones observed in the previous study on a polycrystalline sample.\cite{Oyama} According to the study,\cite{Oyama} the anomaly at 15 K arises from the emergence of a long-range antiferromagnetic ordering of Nd$^{3+}$, while a short-range ordering appears below 36 K. The actual long-range ordering temperature ($T_N$) can be lower than 15 K which was assigned from the  $\chi$($T$) peak and the onset of increase of the specific heat. Indeed, the derivative of $\chi$ ($d\chi/dT$) shows a peak at $T$ = 5.9 K, which is also close to the temperature of the specific heat maximum (shown later). Therefore, we assigned $T_N$ = 5.9 K in this paper. The other anomaly at 38 K is due to the spin-reorientation transition of the Fe$^{3+}$ magnetic moments. The spin configuration of the Fe$^{3+}$ above and below the transition has been determined by M\"ossbauer spectroscopy and powder neutron diffraction measurements in Ref.~\onlinecite{Oyama}, where the spin-reorientation was attributed to the onset of the short-range ordering of Nd$^{3+}$. In contrast, no anomaly was observed with the magnetic field along the $a$-axis, which indicates a strong anisotropy of the spin configuration. In the case of a high magnetic field of $H$ = 9 T (Fig.~3(b)), the  $\chi$($T$) with the magnetic field along the $c$-axis shows an abrupt jump at 5.6 K while cooling down and saturates at lower $T$. This feature can be understood in terms of a transition of the Nd$^{3+}$ moments from the antiferromagnetic state to a spin-polarized state. We also observed that the spin-reorientation transition temperature ($T_{\textrm{SR}}$) of the Fe$^{3+}$ moments shifts to lower temperatures with magnetic fields (from 38 K at 0.1 T to 29 K at 9 T).

\begin{figure}[tbp]
\linespread{1}
\par
\includegraphics[width=3.4in]{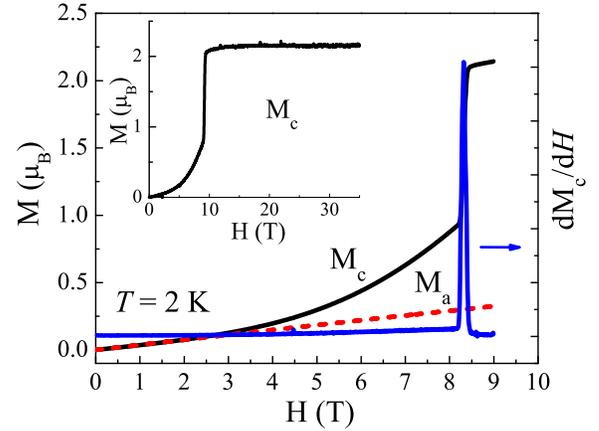}
\caption{  Magnetization versus magnetic field along the $c$-axis (solid) and the $a$-axis (dashed), and $dM_c/dH$ (right axis) versus magnetic field along the $c$-axis at 2 K. The inset shows the magnetization versus magnetic field along the $c$-axis between 0 T and 35 T.}
\end{figure}

Fig. 4 shows the magnetization along the $c$ and $a$-axis as a function of the magnetic field at 2 K. The magnetization grows monotonically as a function of field along both the $c$ and $a$-axis. However, the magnetization for $H\parallel c$ starts to increase faster than for $H\parallel a$ for $H >$ 3 T, shows an abrupt jump at 8.3 T, and finally saturates with a magnetic moment of about 2.15 $\mu_B$ above 10 T as shown in the inset of Fig.~4. The saturation moment is less than the theoretical value of free Nd$^{3+}$ moments (3.27 $\mu_B$), which is probably due to the crystalline electric field effect. The rapid increase of the magnetization followed by the immediate saturation suggests that the spin-flop transition (spin-flops from the antiferromagnetic easy axis ($c$-axis) to the $ab$-plane) and the complete spin polarization (spins are completely parallel to the $c$-axis) occur in a very narrow field range. The field dependence is also consistent with the $\chi$($T$) shown in Fig.~3(b). In contrast, the magnetization along the $a$-axis shows a steady increase as expected for the magnetization process of antiferromagnets when the magnetic field is applied perpendicular to the easy axis.

\begin{figure}[tbp]
\linespread{1}
\par
\includegraphics[width=3.4in]{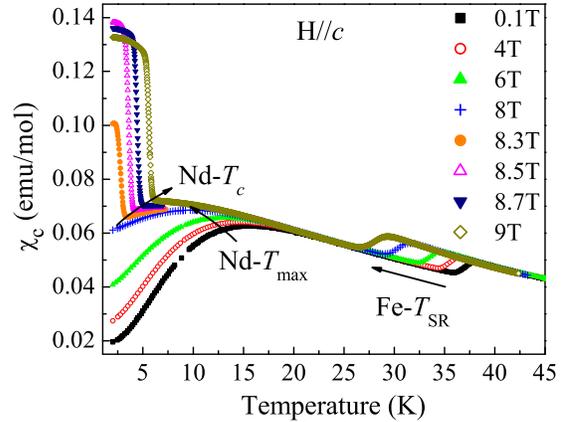}
\caption{  $\chi$($T$) in $H$ fields between 0 T and 9 T along the $c$-axis. The spin-flop transition $T_{c}$ of the Nd$^{3+}$ spins at lower $T$ and the spin reorientation transition $T_{\textrm{SR}}$ of the Fe$^{3+}$ spins at higher $T$ are seen.}
\par
\end{figure}

We have measured the temperature dependence of the magnetic susceptibility in greater detail as shown in Fig.~5. Note that the temperature of the maximum $\chi_c$ for the Nd$^{3+}$ ($T_{\textrm{max}}$) and $T_{\textrm{SR}}$ for the Fe$^{3+}$ decrease with increasing magnetic field, which is a usual behavior of antiferromagnets. On the other hand, the temperature for the complete spin polarization of Nd$^{3+}$ (denoted as $T_c$) increases with fields, which is a feature of ferromagnetic transitions as well as of spin-flop transitions of antiferromagnets.

\begin{figure}[tbp]
\linespread{1}
\par
\includegraphics[width=3.4in]{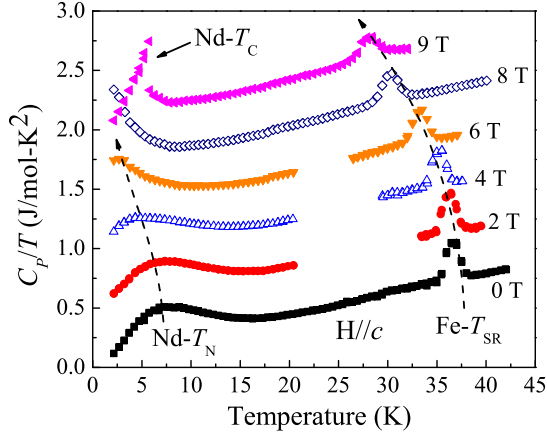}
\caption{  Temperature dependence of $C_p/T$ for $H$ fields between 0 T and 9 T along the $c$-axis. The spin-flop transition ($T_c$) and the N\'eel ordering transition ($T_N$) of the Nd$^{3+}$ spins at lower $T$ and the spin reorientation transition $T_{\textrm{SR}}$ of the Fe$^{3+}$ spins at higher $T$ are seen (as in Fig.~5).}
\par
\end{figure}

The specific heat measurements are consistent with the features observed in the $\chi$($T$) data. The ratio of the specific heat over the temperature, $C_p(T)/T$, is plotted as a function of $T $ in Fig.~6. The broad maximum at low temperatures for fields below 8 T is associated to the long-range antiferromagnetic transition of the Nd$^{3+}$ moments and its position is in agreement with the $\chi$($T$) data (Fig.~3(a) and Fig.~5). The absence of a sharp feature suggests that there is a gradual onset of the Nd$^{3+}$ spin order. As for $\chi_c$, $T_N$ shifts to lower temperatures as the magnetic field increases. However, above 8 T, the spins of Nd$^{3+}$ moments flop to a spin polarized state and the specific heat shows a sharp transition in the 9 T curve, which is to be distinguished from the broad peak at the N\'eel temperature for $H$ = 8 T and below. In addition, the high temperature anomaly at about 28 - 37 K due to the spin reorientation transition of the Fe$^{3+}$ moments also shows a similar behavior as in the  $\chi$($T$) data. Again, $T_{\textrm{SR}}$ shifts to lower temperatures as the magnetic field increases.

\begin{figure}[tbp]
\includegraphics[width=3.0in]{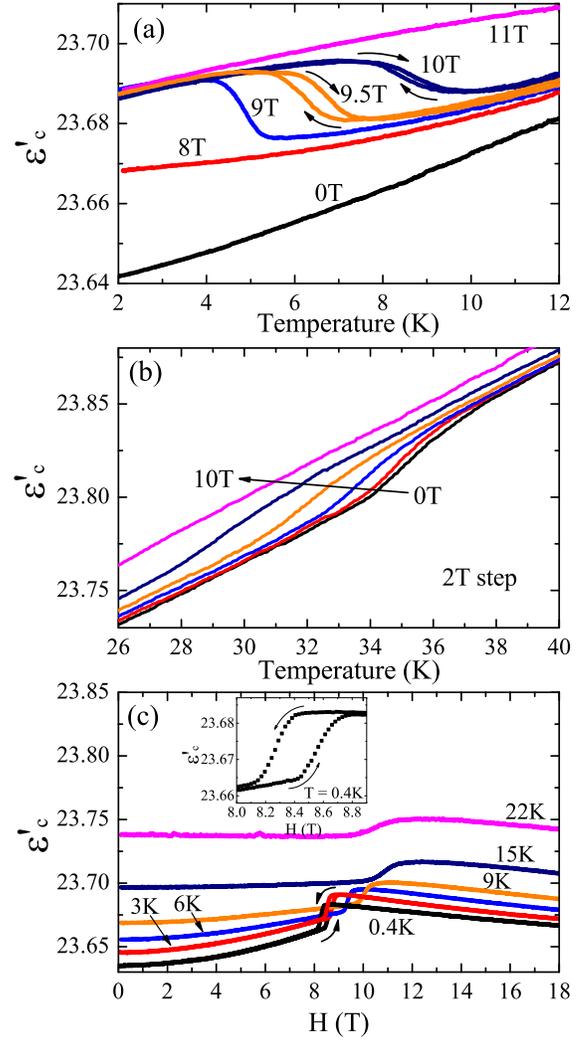}
\caption{  The real part of the dielectric permittivity with both the magnetic and the electric fields applied along the $c$-axis. In (a) and (b) the temperature dependence for various magnetic fields and in (c) the magnetic field dependence for various temperatures are shown.}
\end{figure}

\begin{figure}[tbp]
\linespread{1}
\par
\includegraphics[width=3.0in]{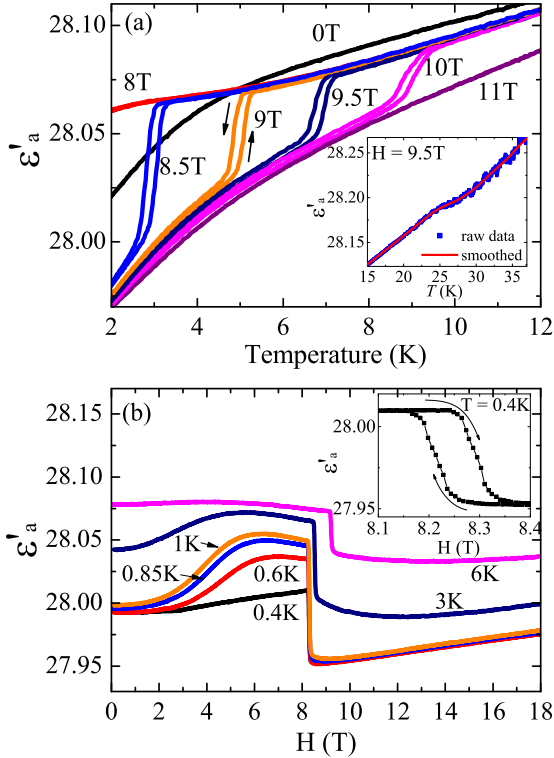}
\par
\caption{  (a) Temperature dependence and (b) magnetic field dependence of the real part of the dielectric permittivity with the magnetic field applied along the $c$-axis and the electric field along the $a$-axis.}
\end{figure}

The temperature and magnetic field dependencies of the real part of the dielectric permittivity ($\varepsilon^\prime$) along the $c$ ($\varepsilon_c^\prime$) and $a$-axis ($\varepsilon_a^\prime$) are shown in Fig.~7 and 8, respectively. The dielectric permittivity data show features associated with the spin reorientation of Fe$^{3+}$ and spin-flop transitions of Nd$^{3+}$, while no anomalies are observed around the $T_N$ of Nd$^{3+}$. The spin-flop transition is most visible in the dielectric permittivity data as a sharp increase (decrease) of $\varepsilon_c^\prime$ ($\varepsilon_a^\prime$). The change of $\varepsilon^\prime$ at the spin-flop transition is more significant for the $a$-axis. For example, when the dielectric permittivity change ($\Delta\varepsilon^\prime$) is normalized by the value just below the spin-flop transition ($\varepsilon^\prime_0$) (defined as magnetodielectric effect here and after), $\Delta\varepsilon^\prime$/$\varepsilon^\prime_0\vert_{a-\rm{axis}}$ = -0.0020 while $\Delta\varepsilon^\prime$/$\varepsilon^\prime_0\vert_{c-\rm{axis}}$ = +0.00076 at $T$ = 0.4 K. These features show hysteresis behavior as shown in Fig.~7(a) and (c), and in Fig.~8(a) and (b). On the other hand, weaker anomalies appear near the Fe$^{3+}$ reorientation temperature with opposite behaviors (decrease of $\varepsilon_c^\prime$ and increase of $\varepsilon_a^\prime$) (Fig.~7(b) and the inset of Fig.~8(a)). When the magnetic field is applied along the $a$-axis, both $\varepsilon_a^\prime$ and $\varepsilon_c^\prime$ increase monotonically with the field (data not shown here). We also measured the pyroelectric current for $E~\| ~H ~\| ~c$-axis configuration under magnetic fields but could not detect any signal. The field-temperature relation of the dielectric permittivity anomalies is consistent with what is obtained from the magnetization and specific heat measurements, which allowed us to build a phase diagram.

\begin{figure}[tbp]
\linespread{1}
\par
\includegraphics[width=3.4in]{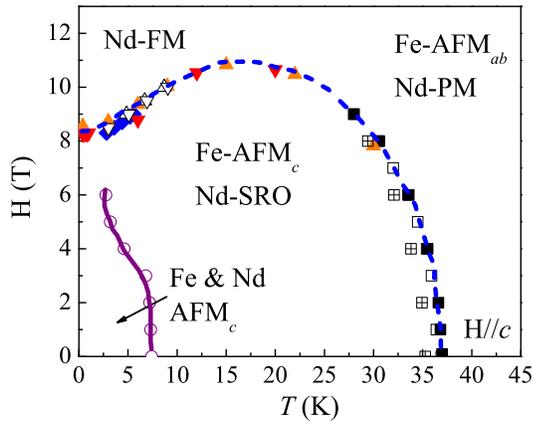}
\par
\caption{  $H - T$ phase diagram for single crystals of SrNdFeO$_4$. The magnetic field is applied along the crystallographic $c$-axis. The solid line represents the second order transition while the dashed line the first order transition. The meaning of each phase can be found in the text. The symbols represent: $\color{black}\blacksquare$ Fe$^{3+}$ $T_{\textrm{SR}}$ from $C_p$, $\color{black}\square$ Fe$^{3+}$ $T_{\textrm{SR}}$ from $\chi_m$, $\color{black}\boxplus $ Fe$^{3+}$ $T_{\textrm{SR}}$ from $\varepsilon'_c$ ($T$ sweep), $\color{purple}\bigcirc$ Nd$^{3+}$ $T_N$ from $C_p$, $\color{blue}\diamondsuit$ Nd$^{3+}$ $T_c$ from $\chi_m$, $\color{orange}\blacktriangle$ Nd$^{3+}$ $T_c$ from $\varepsilon'_c$ ($H$ sweep), $\color{black}\triangle$ Nd$^{3+}$ $T_c$ from $\varepsilon'_c$ ($T$ sweep), $\color{red}\blacktriangledown$ Nd$^{3+}$ $T_c$ from $\varepsilon'_a$ ($H$ sweep), $\color{black}\triangledown$ Nd$^{3+}$ $T_c$ from $\varepsilon'_a$ ($T$ sweep).}
\end{figure}

The magnetic field vs.~temperature phase boundaries of SrNdFeO$_4$ below 40 K are displayed in Fig.~9 by putting all the results from the $\chi$($T$), the specific heat and the dielectric permittivity together. The M\"ossbauer data of Ref.~\onlinecite{Oyama} indicate that the Fe$^{3+}$ moments disorder at about 360 K with a second order transition, which is not shown in this phase diagram. Below $T$ = 8 K and $H$ = 8 T, the phase diagram shows a boundary for a second order phase transition from the short-range (Nd-SRO) to the long-range ordered state along the $c$-axis (Nd-AFM$_c$) of the Nd$^{3+}$ moments. Above 38 K, the Fe$^{3+}$ moments are antiferromagnetically aligned in the $ab$-plane (Fe-AFM$_{ab}$).  However, below 38 K, the Fe$^{3+}$ spins reorient to align with the magnetic field applied along the $c$-axis (Fe-AFM$_{c}$). The $T_{\textrm{SR}}$ decreases with magnetic field and, at low temperature and high magnetic fields, it displays a re-entrance at a constant magnetic field. At low temperature and high magnetic fields, the Nd$^{3+}$ moments flop and form a spin-polarized state. The spin-flop of the Nd$^{3+}$ moments and the reorientation of the Fe$^{3+}$ moments are first order transitions merging into a single boundary. The spin-polarized phase of the Nd$^{3+}$ spins (Nd-FM) at high fields and low $T$ appears to have long-range ferromagnetic order, as evidenced by the jump in the magnetization (Fig.~4), the  $\chi$($T$) (Fig.~5) and the specific heat (Fig.~6). It is remarkable that the dielectric permittivity measurements clearly display all the first order transitions, which suggests strong magnetoelectric and/or magnetoelastic couplings in SrNdFeO$_4$.

The magnetization and the M\"ossbauer spectra for the isostructural sister compound SrLaFeO$_4$ have been measured in Ref.~\onlinecite{Kawanaka}. In this compound the magnetic Nd$^{3+}$ ions are replaced by the nonmagnetic La$^{3+}$ ions. As a consequence only the spin reorientation boundary of the Fe$^{3+}$ moments is observed, which is in qualitative agreement with our observations, including the re-entrant behavior at low temperatures. The transition displays hysteresis in the magnetization, indicating its first order nature. For SrLaFeO$_4$, the required field is smaller than that for SrNdFeO$_4$ and the transition temperature is higher. This indicates that the exchange couplings in SrLaFeO$_4$ are likely to be larger than in SrNdFeO$_4$, but the coercivity is weaker, since the crystalline fields of Nd$^{3+}$ enhance the exchange anisotropy.

\section{Discussion}

\begin{figure*}[tbp]
\linespread{1}
\par
\includegraphics[width=6.8in]{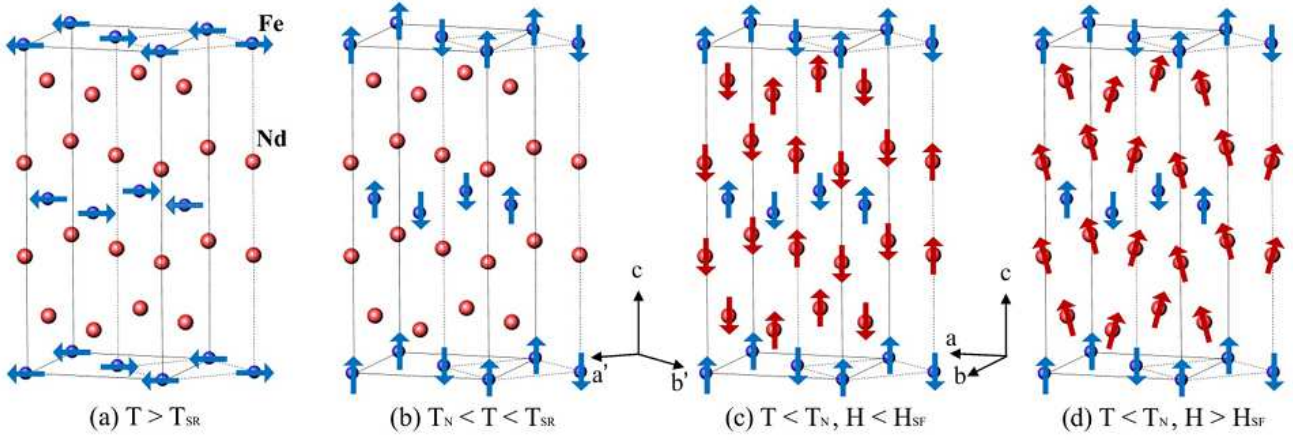}
\par
\caption{  Magnetic structure model of SrNdFeO$_{4}$  for $H$ along the $c$-axis. (a) Above $T_{c}$ the Fe$^{3+}$ spins are ordered antiferromagnetically with the ordered moment lying in the $ab$-plane, while the Nd$^{3+}$ spins are disordered. (b) For $T_N < T < T_{\textrm{SR}}$ the Fe$^{3+}$ spins are reoriented and the antiferromagnetism is along the $c$-axis, while the Nd$^{3+}$ spins remain disordered. (c) For $T$ below $T_N$ and $H < H_{\textrm{SF}}$ (spin-flop transition field), both, the Fe$^{3+}$ and the Nd$^{3+}$ sublattices, are antiferromagnetically ordered with the spins aligned with the $c$-axis. (d) For $T < T_N$ and $H > H_{\textrm{SF}}$, all Nd$^{3+}$ moments are aligned parallel to the field. Red spheres are Nd spins and blue spheres are Fe spins. Diamagnetic ions are omitted.}
\end{figure*}

In this paper we discovered a spin-flop transition of the Nd$^{3+}$ moments at $T_{c}$ and the strongly correlated magnetic ordering of the Nd$^{3+}$ and the Fe$^{3+}$ moments, which manifests itself in the formation of a single phase boundary from two different magnetic transitions. There is a strong evidence of a magnetoelectric and/or magnetoelastic coupling, since the first order magnetic transitions can be clearly seen in the dielectric permittivity. Below we discuss the possible spin structure of the compound and the possible origins for the dielectric anomaly.

\subsection{Spin structure}

In Fig.~10, we schematically present the possible spin structure for the four ordered phases of SrNdFeO$_4$. At high temperatures ($T \! > \! T_{\textrm{SR}}$) only the Fe$^{3+}$ moments are ordered. The long range order is that of a simple antiferromagnet with the moments contained in the $ab$-plane (see Fig.~10(a)). As the temperature is lowered, the Fe$^{3+}$ moments reorient below $T_{\textrm{SR}}$ in a magnetic field parallel to the $c$-axis. At the same time, a short range order of Nd$^{3+}$ emerges as the result of antiferromagnetic interaction between Fe$^{3+}$ and Nd$^{3+}$. The new order of the Fe$^{3+}$ spins is still antiferromagnetic, but the ordered moments are now oriented along the $c$-axis (see Fig. 10(b)). Below $T_N$, the Nd$^{3+}$ spins participate in the antiferromagnetic long-range order (see Fig.~10(c)), with the magnetic moments of both sublattices (Fe$^{3+}$ and Nd$^{3+}$ spins) being aligned with the $c$-axis (see Fig.~10(c)). If the magnetic field is increased beyond $H_{\textrm{SF}}$, the Nd$^{3+}$ moments flop into complete polarization (see Fig.~10(d)), as evidenced by the saturation of the magnetization in the inset of Fig.~4. All the transitions displayed in the phase diagram (Fig.~9), except $T_N$, appear to be of first order. The proposed spin structures are consistent with the findings and interpretations of Refs.~\onlinecite{Oyama} and \onlinecite{Kawanaka}.

The proposed changes of the spin configurations at the transitions explain all the presented experimental results, except for the dielectric permittivity, which requires additional theoretical hypothesis to connect the magnetic transitions with displacements of charges in the compound.

\subsection{Magnetodielectric effect}

In magnetic multiferroics, the dielectric permittivity anomalies under magnetic fields can be clearly seen as the sample undergoes transitions between paralelectric and ferroelectric phases. But in many systems where the multiferroicity is not clearly present, the magnetodielectric effect has been also observed and was contributed to various mechanisms such as linear\cite{Mufti,Hwang} or biquadratic\cite{Kimura_BiMnO3,Yang} magnetoeletric (ME) effect or magnetostriction.\cite{Yen} In SrNdFeO$_4$, we could not detect any pyroelectric current signal around the transitions, so we exclude the possibility of magnetic multiferroicity as a source of the dielectric anomaly.

The ME effect arises due to electric dipolar fields induced due to small displacements of ions by the magnetic field or at a magnetic transition. These electric dipolar fields reflect in changes of the dielectric function. In general, the free energy of a material under external electric ($E$) and magnetic ($H$) fields can be expressed as\cite{LawesSSC},
\begin{eqnarray}
F & = & (1/2\varepsilon_0) P^2 - PE - \alpha PM \nonumber \\
&& +\: \beta PM^2 +\gamma P^2 M +\delta P^2 M^2
\end{eqnarray}
where $\varepsilon_0$ is the bare dielectric permittivity and $\alpha$, $\beta$ and $\gamma$, $\delta$ are coupling constants and $P$ and $M$ are the polarization and the magnetization respectively. Here the free energy is expanded in terms of $P$ and $M$ for simplicity since the only external field in the present work is the magnetic field applied along the $c$-axis. We also omitted the terms not coupled to $P$, because the magnetodielectric effect can be derived from the second derivative of the free energy with respect to $P$. A full expansion of the free energy in terms of $E$ and $H$ including omitted terms here can be found elsewhere.\cite{Shvartsman,Schmid} The linear ME effect is originated from the third term ($\alpha P M$) while the quadratic (biquadratic) ME term from $\beta PM^2$ and $\gamma P^2 M$ ($\delta P^2 M^2$). Due to the different parity of $P$ and $M$ under time reversal and the inversion operation, the linear and the quadratic ME effect are possible when certain symmetry conditions are satisfied. For example, the linear ME effect survives only when both time reversal and inversion symmetries are broken. On the other hand, the biquadratic ME term can be present without restriction of either crystal or magnetic symmetries.

In an attempt to explain the dielectric permittivity anomaly of SrNdFeO$_4$, we first examined the magnetic point group symmetry implied by the spin structure depicted in Fig.~10 for $T \! < \! T_{SR}$. Assuming the crystal point group symmetry $4/mmm$ is conserved, there are 6 possible magnetic point subgroups, $4/m^\prime m^\prime m^\prime$, $4/m^\prime mm$, $4^\prime/m^\prime m^\prime m$, $4/m m^\prime m^\prime$, $4/m m m$ and $4^\prime/m m m^\prime$. Among these, magnetic point subgroups with time reversal symmetry along the $c$-axis ( $4/m^\prime m^\prime m^\prime$, $4/m^\prime mm$, $4^\prime/m^\prime m^\prime m$) allow a linear magnetoelectric effect, which could give a large magnetodielectric effect at the magnetic transition.\cite{Schmid} In SrNdFeO$_4$, it is obvious that the time reversal symmetry is not conserved along the $c$-axis (spins normal to the (001) planes are antiparallel). In other words, the time reversal symmetry is broken for a mirror operation for the (001) mirror plane and so it is for the (100) mirror plane. On the other hand, the time reversal symmetry is conserved for a mirror operation for the (110) mirror plane and for a 4-fold rotation around the [100] axis. To sum up, $4^\prime/m m m^\prime$ is the possible magnetic point group symmetry of SrNdFeO$_4$.\cite{Graef} According to the symmetry analysis,\cite{Schmid} the lowest order allowed ME coefficient for the $4^\prime/m m m^\prime$ symmetry is the quadratic $\gamma P^2 M$.

Since the linear ME effect is not allowed from the symmetry, the magnetodielectric effect, if any, should be due to the quadratic $\gamma P^2 M$ and/or the biquadratic term. The biquadratic term is symmetry independent and proportional to $\delta P^2 M^2$, where $\delta$ can be either a negative or positive constant. In a phenomenological model suggested by Lawes et al.\cite{Lawes}, $\delta P^2 M^2$ is replaced by $\sum_{q}^{} g(q)P^2 \left \langle  M_q M_{-q}\right \rangle$ (where $\left \langle  M_q M_{-q}\right \rangle$ is the $q$ dependent magnetic correlation function and $g(q)$ is the spin-lattice coupling constant). This model successfully explained the magnetodielectric effect of some ferromagnetic and antiferromagnetic materials. But neither the quadratic or the biquadratic term can explain the sign difference of the magnetodielectric effect between the $c$ and $a$-axis, since the magnetic field was along the $c$-axis for both configurations, where only an increase of the magnetization was observed.

The sign difference of the magnetodielectric effect can be well understood if one considers magnetostriction without invoking the ME effect. In this case, the information we obtain from the capacitance measurement is rather about the geometrical dimensions than about the dielectric permittivity. The capacitance $C$ can be expressed as $\varepsilon^\prime \times (\rm{Area} / \rm{thickness})$ for the parallel capacitor geometry we used in this work. If one considers the magnetostriction alone and assumes the total unit cell volume ($V_{\textrm{cell}}$) and the tetragonal symmetry is conserved, the magnetocapacitance ($\Delta C_{a(c)}/C_{a(c)}$) for the $a$($c$)-axis can be expressed as following,
\begin{eqnarray}
\Delta C_a /C_a & = & \Delta (\varepsilon^\prime \frac{ac}{a}) / (\varepsilon^\prime \frac{ac}{a}) = \Delta c /c\ , \\
\Delta C_c /C_c & = & \Delta (\varepsilon^\prime \frac{a^2}{c}) / (\varepsilon^\prime \frac{a^2}{c}) = -2\Delta c /c
\end{eqnarray}
where $\varepsilon^\prime$ is the magnetic field independent dielectric permittivity and $a$, $c$ are the lattice constants. Since the magnetostriction causes one crystallographic axis to shrink while the other to expand, it is clear that the magnetostriction can explain the signs of the magnetodielectric effect (or more correctly magnetocapacitance in this case), if the crystal shrinks along the $c$-axis ($\delta c/c \leq 0$) at the spin-flop transition. However, it fails to explain why $\Delta C_a /C_a$ (= +0.0020) is more than 2 times larger than $\Delta C_c /C_c$ (= -0.00076) instead of 2 times smaller.

As an alternative possibility, we considered a case where both the quadratic ME effect and the magnetostriction are active at the same time. In this case, the $\varepsilon^\prime$ in Eqs.~(2) and (3) are also magnetic field dependent, hence a ($\Delta \varepsilon ^\prime / \varepsilon^\prime)_{\textrm{ME}}$ term should be added. Then, the above equations can be written as,
\begin{eqnarray}
\Delta C_a /C_a & = & (\Delta \varepsilon ^\prime / \varepsilon^\prime)_{\textrm{ME}} + \Delta c /c\ , \\
\Delta C_c /C_c & = & (\Delta \varepsilon ^\prime / \varepsilon^\prime)_{\textrm{ME}} - 2\Delta c /c
\end{eqnarray}
The ($\Delta \varepsilon ^\prime / \varepsilon^\prime)_{\textrm{ME}}$ term is solely dependent on the magnetization and can be either positive or negative. Then the observed magnetocapacitance behavior (sign and magnitude) can be qualitatively explained when ($\Delta \varepsilon ^\prime / \varepsilon^\prime)_{\textrm{ME}}$ and $\Delta c /c$ are in a certain range. Note that ($\Delta \varepsilon ^\prime / \varepsilon^\prime)_{\textrm{ME}}$ should always be positive in any case to satisfy the signs of the observed magnetocapacitance. Quantitatively, when we use the observed magnetocapacitance values, we obtain ($\Delta \varepsilon ^\prime / \varepsilon^\prime)_{\textrm{ME}}$ = 0.00108 and $\Delta c /c$ = 0.00092. These values are rather large (small) for typical magnetostriction (non-linear ME) effect\cite{Handley}, which calls for further refinements experimentally and theoretically. For example, an independent magnetostriction measurement could be used to separate the two effects. The thermal expansion measurement would be also useful to examine the hypothesis of conservation of the tetragonal symmetry and the unit cell volume.

The above arguments should also hold for the dielectric permittivity anomalies observed near the spin reorientation transition of Fe$^{3+}$. At least for the ME effect part, we speculate that the dielectric permittivity change is more likely due to the short-range ordering of Nd$^{3+}$ rather than directly from the reorientation of Fe$^{3+}$ moments. The electronic configuration of the $4f$-shell of Nd$^{3+}$ has large spin-orbit coupling and considerable crystalline field splittings. Due to the spin-orbit coupling, the spin of the configuration is not a good quantum number and is intimately coupled to the orbital angular momentum. The orbital angular momentum can couple to the lattice and hence induce small displacements of the Nd$^{3+}$ ions, which, on the other hand, could generate small dielectric dipoles. This mechanism, however, does not couple to the Fe$^{3+}$ configuration, which is an $S$-state without net orbital content.

Finally, the absence of the dielectric permittivity anomaly around the long-range ordering of Nd$^{3+}$ suggests that neither the ME effect nor the magnetostriction is associated with the transition. Being a local probe sensitive to the electric dipolar fields of ions, the dielectric permittivity can be sensitive to a short- or long-range ordering, while it is insensitive to a transition between short- and long-range orderings.

In summary, we have investigated the magnetic phase diagram of SrNdFeO$_4$ with various experimental techniques. Especially, the intriguing behavior of the magnetocapacitance suggests that the capacitance measurement can be a very sensitive probe for the magnetic phase transitions. When measured on a single crystal sample under different electric/magnetic field configurations, the magnetocapacitance probe also provides a useful opportunity to investigate magnetic and electric couplings.

\begin{acknowledgments}
This work was performed at the National High Magnetic Field Laboratory which is supported by NSF Cooperative Agreement No.~DMR-0654118 and by the State of Florida. P.~S.~is supported by the DOE under Grant No.~DE-FG02-98ER45707. We are grateful to Tim Murphy, Ju-Hyun Park and Glover Jones for their help with experiments carried out in the milikelvin facility of the NHMFL.
\end{acknowledgments}


\begin{thebibliography}{99}

\bibitem{Buschow} K.H.J. Buschow, Rep. Prog. Phys. {\bf 54}, 1123 (1991).
\bibitem{Hummler} K. Hummler and M. F\"ahnle, Phys. Rev. B {\bf 53}, 6 (1996).
\bibitem{Chapon} L.C. Chapon, G.R. Blake, M,H, Gutmann, S. Park, N. Hur, P.G. Radaelli, and S.-W. Cheong, Phys. Rev. Lett. {\bf 93}, 177402 (2004).
\bibitem{Tokunaga} Y. Tokunaga, S. Iguchi, T. Arima, and Y. Tokura, Phys. Rev. Lett. {\bf 101}, 097205 (2008).
\bibitem{Kimura_TbMnO3} T. Kimura, T. Goto, H. Shintani, K. Ishizaka, T. Arima, and Y. Tokura, Nature {\bf 426}, 55 (2003).
\bibitem{Kenzelmann} M. Kenzelmann, A.B. Harris, S. Jonas, C. Broholm, J. Schefer, S.B. Kim, C.L. Zhang, S.-W. Cheong, O.P. Vajk, and J.W. Lynn, Phys. Rev. Lett. {\bf 95}, 087206 (2005).
\bibitem{Arima} T. Arima, A. Tokunaga, T. Goto, H. Kimura, Y. Noda, and Y. Tokura, Phys. Rev. Lett. {\bf 96} (2006) 097202.
\bibitem{Hur} N. Hur, S. Park, P.A. Sharma, J.S. Ahn, S. Guha, and S.-W. Cheong, Nature, {\bf 429}, 392 (2004).
\bibitem{Oyama} S. Oyama, M. Wakeshima, Y. Hinatsu, and K. Ohoyama, J. Phys.: Condens. Matter. {\bf 16}, 1823 (2004).
\bibitem{Komskii} D.I. Komskii, J. Magn. Magn. Mater. {\bf 306}, 1 (2006).
\bibitem{Eerenstein} W. Eerenstein, N.D. Mathur, and J.F. Scott, Nature {\bf 442}, 17 (2006).
\bibitem{Cheong} S.-W. Cheong and M. Mostovoy, Nature Materials {\bf 6}, 13 (2007).
\bibitem{Pennycook} S.J. Pennycook and L.A. Boatner, Nature {\bf 336}, 565-567 (1988).
\bibitem{Loane} R.F. Loane, P. Xu and J. Silcox, Acta Crystallographica Section A {\bf 47} 267-278 (1991).
\bibitem{Kawanaka} H. Kawanaka, H. Bando, K. Mitsugi, A. Sasahara, and Y. Nishihara, Physica B {\bf 329-333}, 797 (2003).
\bibitem{Mufti} N. Mufti, G.R. Blake, M. Mostovoy, S. Riyadi, A.A. Nugroho, and T.T.M. Palstra, Phys. Rev. B {\bf 83}, 104416 (2011).
\bibitem{Hwang} J. Hwang, E.S. Choi, H.D. Zhou, J. Lu, and P. Schlottmann, Phys. Rev. B {\bf 85}, 024415 (2012).
\bibitem{Kimura_BiMnO3} T. Kimura, S. Kawamoto, I. Yamada, M. Azuma, M. Takano, and Y. Tokura, Phys. Rev. B {\bf 67}, 180401 (2003).
\bibitem{Yang} Y. Yang, J.-M. Liu, H.B. Huang, W.Q. Zou, P. Bao, and Z.G. Liu, Phys. Rev. B {\bf 70}, 132101 (2004).
\bibitem{Yen} F. Yen, B. Lorenz, Y.Y. Sun, C.W. Chu, L.N. Bezmaternykh, and A.N. Vasiliev, Phys. Rev. B {\bf 73}, 054435 (2006).
\bibitem{LawesSSC} G. Lawes, T. Kimura, C.M. Varma, M.A. Subramanian, N. Rogado, R.J. Cava, A.P. Ramirez, Prog. Sol. State Chem. {\bf 37}, 40 (2009).
\bibitem{Shvartsman} V. V. Shvartsman, P. Borisov, W. Kleemann, S. Kamba, T. Katsufuji, Phys. Rev. B {\bf 81}, 064426 (2010).
\bibitem{Schmid} H. Schmid, J. Phys.: Condens. Matter {\bf 20} 434201 (2008).
\bibitem{Graef} M. De Graef, J. of Materials Education {\bf 20}, (3-4) 31 (1998).
\bibitem{Lawes} G. Lawes, A.P. Ramirez, C.M. Varma, and M.A. Subramanian, Phys. Rev. Lett. {\bf 91}, 257208 (2003).
\bibitem{Handley} R.C. O'Handley, {\it Handbook of Magnetism and Advanced Magnetic Materials} vol. 4, edited by H. Kronmueller and S. Parkin (Wiley, New York, 2007).

\end{thebibliography}
\end{document}